\documentclass[a4paper,12pt]{article}
\usepackage[utf8]{inputenc}
\usepackage[english]{babel}
\usepackage{blindtext}
\usepackage{microtype}
\usepackage{cite}
\usepackage{titlesec}
\usepackage{graphicx}
\usepackage{multirow}
\usepackage{wrapfig}
\usepackage[nottoc,numbib]{tocbibind}
\usepackage{subcaption}
\usepackage{float}
\usepackage{amsmath}
\usepackage{index}
\makeindex

\begin{document}
\pagenumbering{gobble}

\begin{flushright}
 September 8, 2022
\end{flushright}

\begin{center}
    
\end{center}

\begin{center}
\textbf{NuSmear: Fast Simulation of Energy Smearing and Angular Smearing for Neutrino-Nucleon Scattering Events in the GENIE Event Generator}
\end{center}

\begin{center}
    
\end{center}

\begin{center}
Ishaan Vohra
\\
\emph{Phillips Exeter Academy}
\\
\emph{Exeter, NH, USA}
\end{center}

\begin{center}
    
\end{center}

Complete Monte Carlo (MC) simulation of a neutrino experiment typically involves the lengthy and CPU-intensive process of integrating models of incoming neutrino fluxes, event generation, and detector setup. We describe a fast, geometry-independent, GENIE-based system known as NuSmear which provides preliminary simulation of energy smearing and angular smearing for neutrino-nucleon interactions. We discuss NuSmear's simulation methodology, explaining its calculation of reconstructed values through its model-based resolution computation, smearing distributions, and particle detection dependencies. We go on to validate NuSmear's performance as a fast simulation system through a series of internal and external comparisons testing its predictive accuracy and model-adherence. Finally, we briefly explore the potential for future user-customization in NuSmear.

\begin{center}
    
\end{center}

\begin{center}
PRESENTED AT
\end{center}

\begin{center}
Fourteenth Conference on the Intersections \\ of Particle and Nuclear Physics \\ Lake Buena Vista, FL, USA \\ August 29 – September 4, 2022
\end{center}

\pagebreak

\section{Introduction}
\pagenumbering{arabic}
\setcounter{page}{2}

In the realm of modern neutrino physics, the ability of Monte Carlo methods to provide detailed simulations of neutrino-nucleon interactions plays an essential role in both data analysis and the planning of future experiments~\cite{seymour2015monte,tanabashi2018particle, bozzi}. For the sake of generating accurate predictions, however, complete Monte Carlo simulation of a neutrino experiment often requires the meticulous modeling of incoming neutrino fluxes, event generation, and detector setup, resulting in necessarily long computing times and high CPU loads to generate a statistically significant sample of events~\cite{IASELLI1986488,araz2021simplified}. 

In recent years, the solution to this problem has appeared in the form of fast Monte Carlo methods, which provide preliminary simulations of experiments centered around balancing the needs of predictive accuracy and computational speed~\cite{wlodek,siddi2018development}. In the area of detector response smearing simulations, various existing tools have found the comfortable middle ground between these two requirements, ranging from multipurpose collider frameworks such as DELPHES to experiment-specific frameworks such as EIC-Smear and the ATLAS Fast Track Simulation Project~\cite{aschenauer2013eic, accardi2016electron,delaere2013delphes,koch2019response,buckley2020fast,hamilton2011atlas}. Within the neutrino physics community, however, while there are tools offering fast simulations for specific experimental setups, few dedicated systems currently exist to provide rapid preliminary smearing simulations for generic neutrino-nucleon scattering events~\cite{huber2005simulation}.

In this work, we propose NuSmear – a novel software system providing fast, generic, geometry-independent simulations of energy smearing and angular smearing via parameterized model-based presets. For ease of user integration, NuSmear\footnote{NuSmear's code is publicly available at https://github.com/GENIE-MC/Generator/tree/master/src/contrib/ivohra/NuSmear} is written as a contribution package built directly onto the GENIE Monte Carlo event generator, which is used by the majority of neutrino experiments worldwide to simulate interactions between all flavors of neutrinos and nuclear targets within the MeV to PeV energy scales~\cite{andreopoulos2010genie}. 

For a given input data set of neutrino-nucleon scattering events, NuSmear performs smearing in four major steps:

\begin{enumerate}
    \item Smearing model selection 
    \item Computation of resolution
    \item Application of smearing distribution
    \item Consideration of particle detection dependency (energy smearing operation only)
\end{enumerate}

First, the user selects one of NuSmear's two current smearing model presets to perform either the energy smearing or the angular smearing operation. The first smearing model is derived from the Deep Underground Neutrino Experiment's Conceptual Design Report (DUNE-CDR) – it incorporates energy resolution functions, angular resolution values, and particle detection dependencies simulating the preliminary far detector response at the future DUNE experiment~\cite{acciarri2016long,abi2020deep, alion2016experiment}. The second smearing model is the NuSmear Default model, which incorporates more straightforward energy resolution functions, angular resolution values, and particle detection dependencies. Once a model is chosen, particles are indexed by type and assigned to one of 72 energy resolution functions or angular resolution values, from which a resolution\footnote{Unlike the common usage of the term \emph{resolution}, in the context of this paper, an increase in resolution value corresponds to a greater degree of smearing, and vice versa.} is computed or obtained. Subsequently, according to the chosen smearing parameter of energy or angle, either a log-normal or Gaussian distribution is generated and applied to the true value of the smearing parameter, producing a reconstructed value. In the case of angular smearing, further dependencies are considered negligible and the reconstructed value is simply returned to the user as an output. In the case of energy smearing, however, further particle detection dependencies may be considered in reference to the type of particle and the chosen smearing model. In the DUNE-CDR model, for example, a neutron with momentum below a 1 GeV/c threshold has a non-zero probability of entirely escaping detection and returning zero reconstructed energy. Once consideration of particle detection dependencies has been completed, the final reconstructed value of the smearing parameter is returned to the user as an output. Figure 1 provides a flowchart representation of NuSmear's complete simulation process.

\begin{figure}[H]
\includegraphics[width = 220pt]{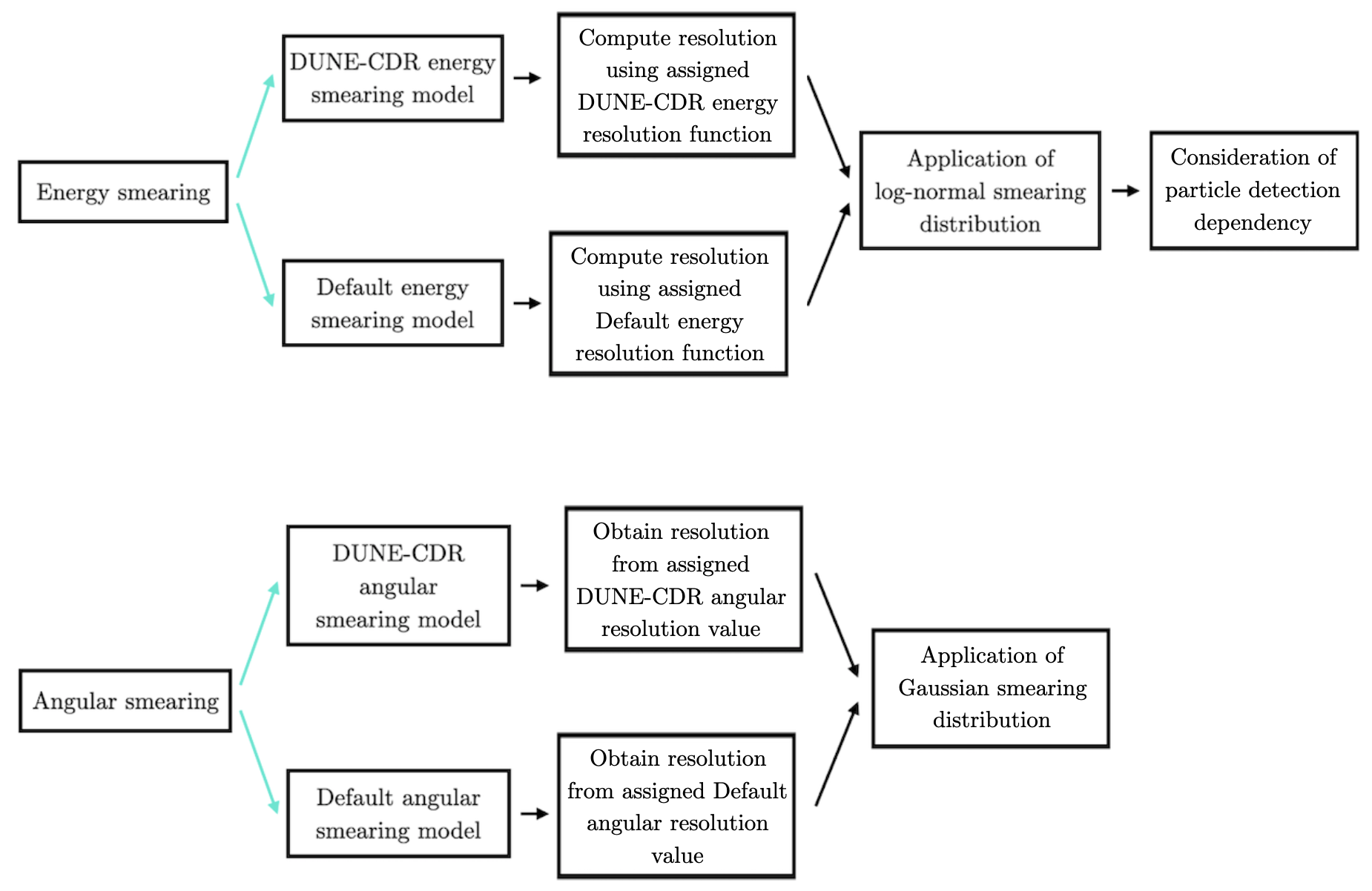}
\centering
\caption{Flowchart diagram depicting NuSmear's smearing simulation process. The turquoise arrows represent decisions made by the user.}
\end{figure}

\section{Simulation methodology}
\vspace*{-3mm}
\subsection{Computation of resolution}
\vspace*{-1mm}
\subsubsection{Energy resolution}

NuSmear's DUNE-CDR energy resolution functions include up to three particle parameters to compute an energy resolution: total energy, kinetic energy, and magnitude of momentum. Kinetic energy and magnitude of momentum are compared to a variety of particle-specific numerical thresholds, returning an energy resolution that is either a constant value or dependent on the total energy~\cite{abi2020deep, alion2016experiment}. Moreover, if a particle does not pass a function's minimum kinetic energy threshold, it is considered undetected, and NuSmear will return zero reconstructed energy to the user. As NuSmear is inherently a geometry-independent smearing system, all DUNE-CDR geometric resolution dependencies are omitted and approximated with numerical values. Table 1 summarizes the calculations performed by NuSmear's DUNE-CDR energy resolution functions.

\begin{table}[H]
\begin{center}
\renewcommand{\arraystretch}{1}
\begin{tabular}{ |c|c|c| } 

 \hline
 Particle type & Function calculations & Omitted dependencies \\ 
 \hline
 \hline
 \multirow{3}{5em}{$\pi^{\pm}$} & if KE $\ge$ 100 MeV, return 15\% & track length,\\ 
&  & showering,\\ 
&  & contained/exiting track\\
\hline
\multirow{2}{5em}{$\gamma$, $e^{\pm}$} & if KE $\ge$ 30 MeV, return & \\ 
& 2\% $\oplus$ 15\%/$\sqrt{E}$[GeV] & \\
\hline
\multirow{3}{5em}{$p$} & if KE $\ge$ 50 MeV, return & \\ 
 & 10\% if $|$p$|$ $<$ 400 MeV/c, else  & \\
  & return 5\% $\oplus$ 30\%/$\sqrt{E}$[GeV] & \\
 \hline
\multirow{2}{5em}{$\mu^{\pm}$} & if KE $\ge$ 30 MeV, return 15\% & track length,\\ 
&  & contained/exiting track\\
\hline
\multirow{2}{5em}{$n$} & if KE $\ge$ 50 MeV, & \\
& return 40\%/$\sqrt{E}$[GeV] &  \\
\hline
\multirow{2}{5em}{other} & if KE $\ge$ 50 MeV, return & \\ 
& 5\% $\oplus$ 5\%/$\sqrt{E}$[GeV] & \\
\hline
\end{tabular}
\caption{Summary of the calculations performed by NuSmear's energy resolution functions in the DUNE-CDR model. $E$, KE, and $|$p$|$ denote total energy, kinetic energy, and magnitude of momentum respectively.}
\end{center}
\end{table}
\vspace*{-5mm}
NuSmear's Default model energy resolution functions involve simpler calculations than that of the DUNE-CDR model. A single comparison is made to check if the particle kinetic energy passes a standard numerical threshold of 50 MeV. If not, the particle is considered undetected and zero reconstructed energy is returned to the user. If so, the particle is assigned a constant energy resolution value according to its type, as summarized in Table 2.
\vspace*{-2mm}
\begin{table}[H]
\begin{center}
\renewcommand{\arraystretch}{1.15}
\begin{tabular}{ |c|c| } 

 \hline
 Particle type & Energy resolution \\ 
 \hline
 \hline
$\pi^{\pm}$,$\pi^{0}$ & 15\%\\ 
\hline
$K^{\pm}$, $K^{0}$/$\bar{K^{0}}$ & 20\%\\
\hline
$\gamma$ & 30\%\\ 
\hline
$e^{\pm}$ & 40\%\\ 
\hline
$p$  & 40\%\\ 
\hline
$\mu^{\pm}$ & 15\%\\ 
\hline
$n$  & 50\%\\ 
\hline
other  & 30\%\\ 
\hline

\end{tabular}
\vspace*{-1mm}
\caption{Summary of the energy resolution values assigned to each particle type in the Default model.}
\end{center}
\end{table}
\vspace*{-10mm}
\subsubsection{Angular resolution}
\vspace*{-1mm}
To obtain an angular resolution for a given particle, NuSmear assigns the particle to a constant angular resolution value determined purely by particle type. The DUNE-CDR model and Default model angular resolution values are itemized in Table 3 below.

\begin{table}[H]
\begin{center}
\renewcommand{\arraystretch}{1.15}
\begin{tabular}{ |c|c|c| } 

 \hline
\multirow{2}{6em}{Particle type} & \multicolumn{2}{|c|}{Angular resolution} \\ 
\cline{2-3}

& DUNE-CDR model & Default model\\
 \hline
 \hline
$\pi^{\pm}$ & 1$^{\circ}$ & 2$^{\circ}$\\ 
\hline
$\pi^{0}$ & 5$^{\circ}$ & 8$^{\circ}$\\ 
\hline
$K^{\pm}$ & 5$^{\circ}$ & 2$^{\circ}$\\
\hline
$K^{0}/\bar{K^{0}}$ & 5$^{\circ}$ & 8$^{\circ}$\\
\hline
$\gamma$, $e^{\pm}$ & 1$^{\circ}$ & 3$^{\circ}$\\ 
\hline
$p$  & 5$^{\circ}$ & 8$^{\circ}$\\ 
\hline
$\mu^{\pm}$ & 1$^{\circ}$ & 2$^{\circ}$\\ 
\hline
$n$  & 5$^{\circ}$ & 10$^{\circ}$\\ 
\hline
other  & 5$^{\circ}$ & 8$^{\circ}$\\ 
\hline

\end{tabular}
\vspace*{-1mm}
\caption{Summary of the angular resolution values assigned to each particle type in the DUNE-CDR and Default models.}
\end{center}
\end{table}
\subsection{Smearing distributions}
\subsubsection{Energy smearing distribution}

Although particle energies are often smeared using a Gaussian distribution, this poses the problem of how one should exclude negative values of energy~\cite{buckley2020fast}. The common solution of truncating the Gaussian distribution reduces the simulation's accuracy to a real detector at low energies, hence NuSmear makes use of the log-normal distribution instead, as shown in Figure 2~\cite{delaere2013delphes, lyons2020statistical}.

\begin{figure}[H]
\includegraphics[width = 350pt]{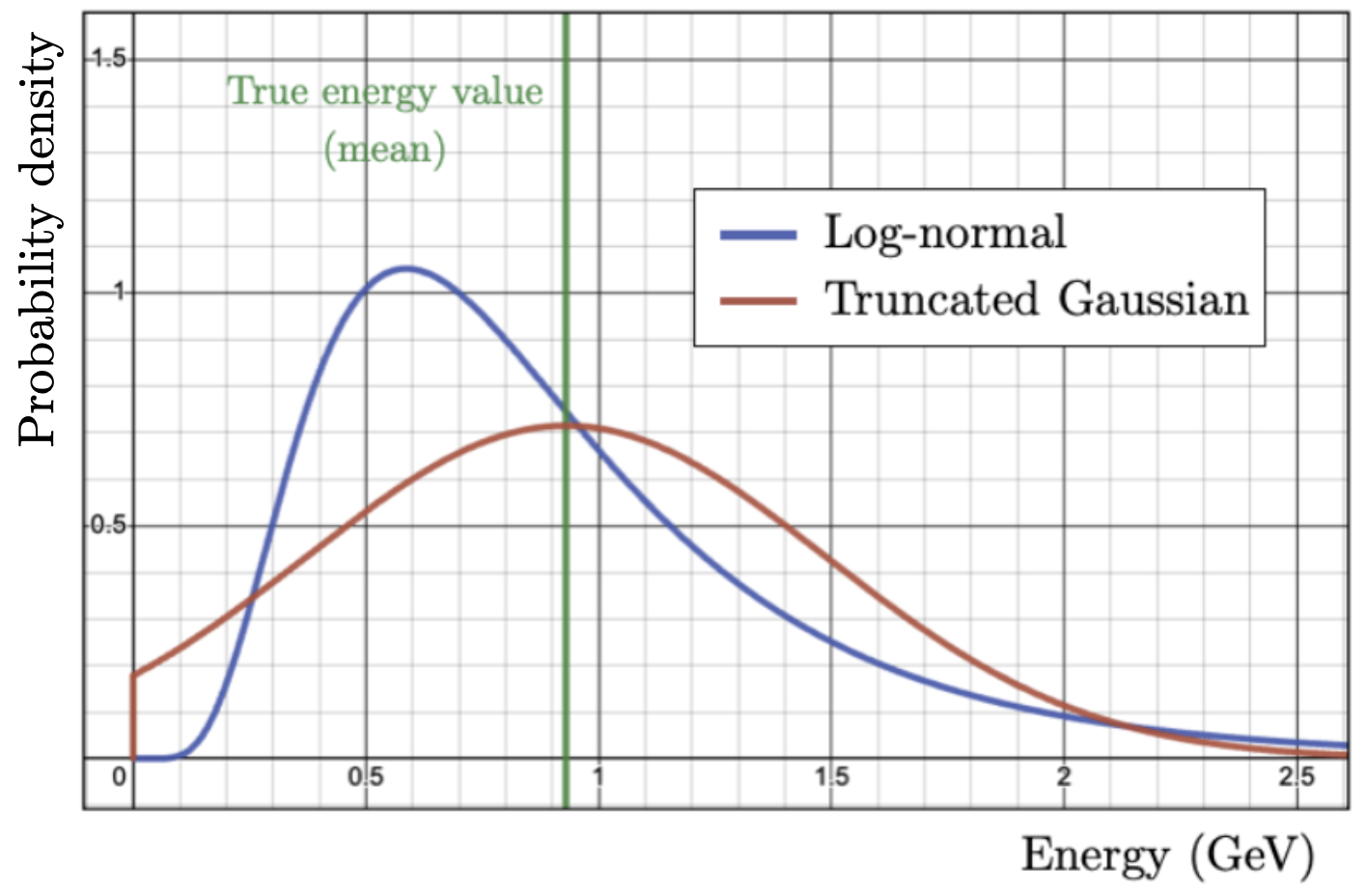}
\centering
\caption{Example log-normal and truncated Gaussian smearing distributions representing the same true energy and resolution values.}
\end{figure}

The log-normal distribution takes the form

\begin{center}
$f(x) = {\frac {1}{x\sigma {\sqrt {2\pi }}}}\ \exp \left(-{\frac {\left(\ln \left(x\right)-\mu \right)^{2}}{2\sigma ^{2}}}\right)$,
\end{center}

in which the distribution parameters $\mu$ and $\sigma$ are given in terms of the mean and variance of the smearing distribution, $m$ and $Var[X]$, by

\begin{center}
$\mu = \ln{\left(\frac{m^2}{\sqrt{Var[X] + m^2}}\right)}$,
\end{center}

\begin{center}
$\sigma ^2 = \ln{\left(1+ \frac{Var[X]}{m^2}\right)}$.
\end{center}

Furthermore, $m$ and $Var[X]$ are physically related to the particle's true energy and calculated energy resolution, $E_{true}$ and $R_E$, via

\begin{center}
$m = E_{true}$,
\end{center}

\begin{center}
$Var[X] = (R_E E_{true})^2$.
\end{center}

Once a particle's smearing distribution is created, NuSmear makes use of the Mersenne Twister pseudo-random number generator (PRNG) to generate the reconstructed energy according to the distribution~\cite{matsumoto1998mersenne}.

\subsubsection{Angular smearing distribution}

In NuSmear, a particle's true outgoing angle with respect to the incident neutrino is represented by $\theta$, as illustrated in Figure 3.

\begin{figure}[H]
\includegraphics[width = 250pt]{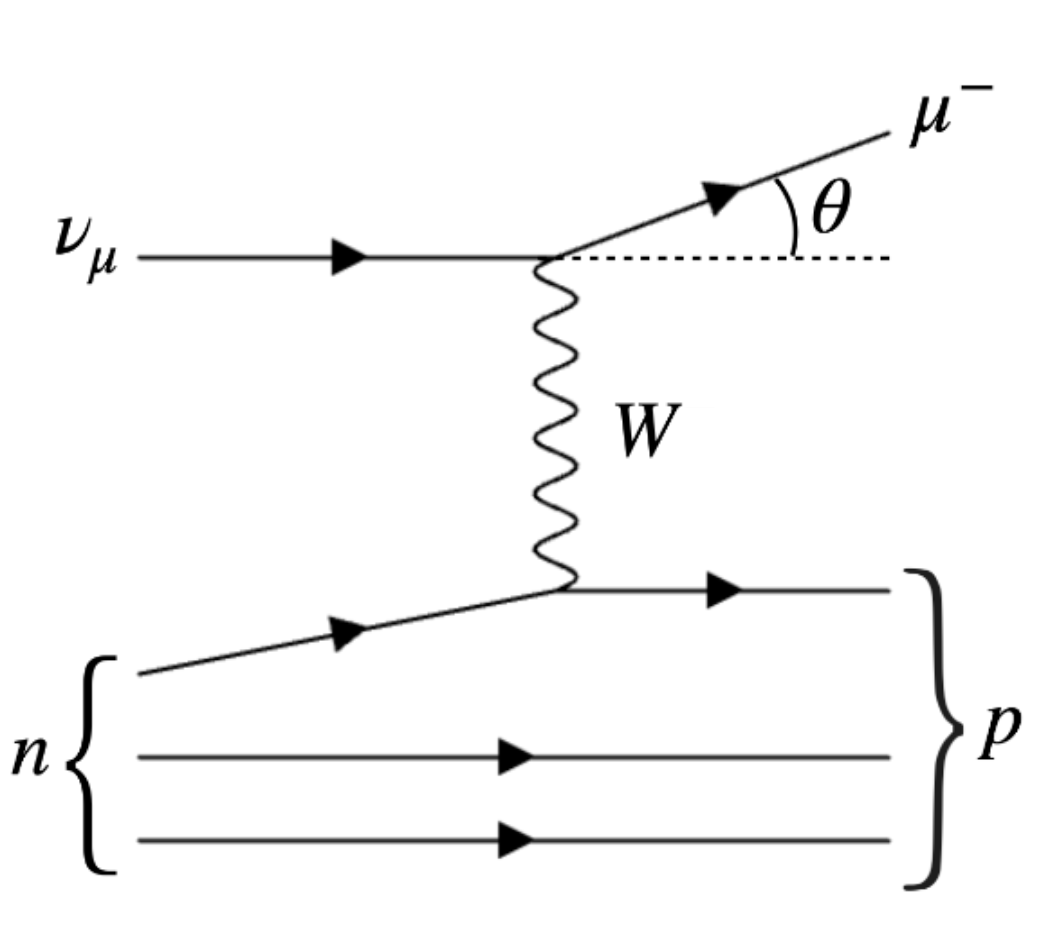}
\centering
\caption{Example charged-current (CC) interaction between an incident muon neutrino and target neutron, producing an outgoing muon at angle $\theta$ with respect to the incident neutrino.}
\end{figure}

For each particle, angular smearing is performed using a Gaussian distribution with a mean set to the true angle, $\theta$, and a standard deviation set to the particle's angular resolution, ${R_A}$, as exemplified in Figure 4.

\begin{figure}[H]
\includegraphics[width = 325pt]{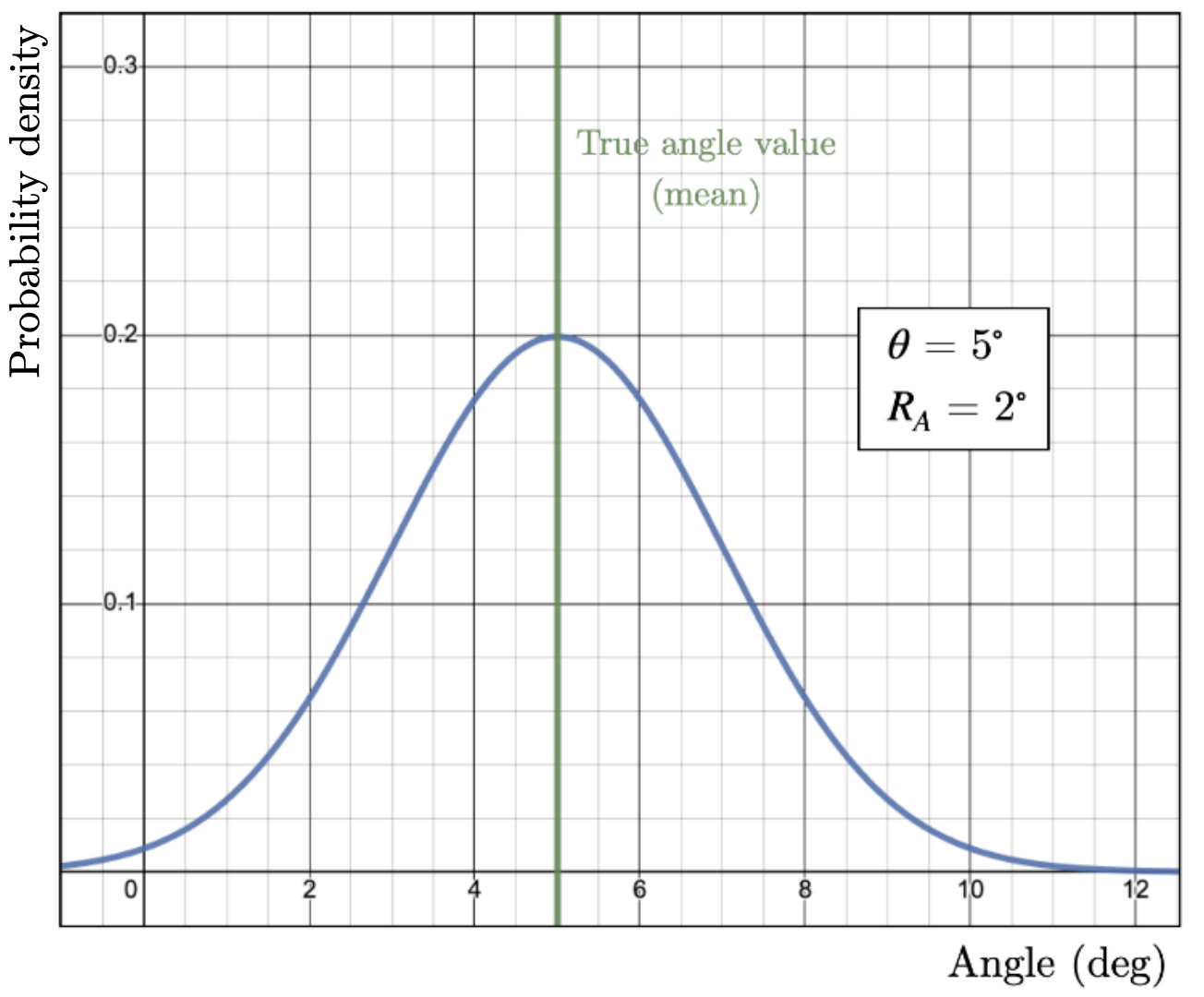}
\centering
\caption{Example Gaussian angular smearing distribution with a true angle value of 5$^{\circ}$ and an angular resolution of 2$^{\circ}$.}
\end{figure}

Alike NuSmear's application of energy smearing distributions, once a particle's angular smearing distribution is created, the Mersenne Twister is used to generate the reconstructed angle according to the distribution~\cite{matsumoto1998mersenne}.

\subsection{Particle detection dependency}

In addition the minimum kinetic energy thresholds incorporated into NuSmear's energy resolution functions, further particle detection dependencies are implemented into the energy smearing operation in order to more accurately simulate unobserved particles in real detectors. Particle detection dependencies are present in both the DUNE-CDR and Default energy smearing models.

Within NuSmear's DUNE-CDR energy smearing model, neutrons with momentum less than 1 GeV/c have a 10\% probability of escaping detection, and for neutrons that are detected, only 60\% of the energy generated by the smearing distribution is returned to the user as the final reconstructed energy~\cite{alion2016experiment}.

In the Default energy smearing model, photons and neutrons have a 50\% probability of escaping detection and returning zero reconstructed energy.

\section{Validation of smearing performance}

To demonstrate the efficacy of NuSmear in providing fast smearing simulations which both maintain predictive accuracy and adhere to their input models, we generate sets of neutrino-nucleon scattering events and perform a series of comparisons between multiple NuSmear simulations, as well as between NuSmear and independent complete Monte Carlo simulations~\cite{agafonova2018final,abe2020measurement}. Within each comparison, we construct and juxtapose multiple smearing matrices, analyzing each simulation through distributions representing the reconstructed smearing parameter in relation to the true smearing parameter.

\subsection{Energy smearing}

\subsubsection{Complete MC comparison}

Generating events using an incoming electron neutrino flux derived from the the OPERA experiment~\cite{acquafredda2009opera}, we evaluate NuSmear's reproduction of the OPERA detector's energy smearing response by constructing a neutrino energy smearing matrix and comparing it to the smearing matrix constructed by the OPERA Collaboration's complete Monte Carlo simulation\cite{agafonova2018final}. Both matrices (see Figure 5 below) are filtered to contain only events involving charged-current (CC) interactions.

\begin{figure}[H]
\includegraphics[width = \textwidth]{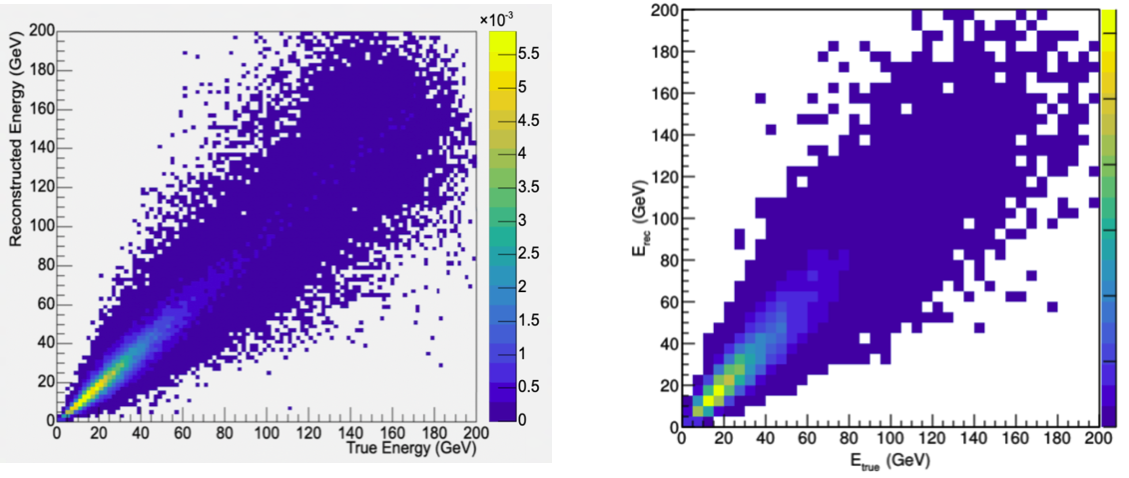}
\centering
\caption{NuSmear Default (left) compared to complete Monte Carlo simulation (right) electron neutrino CC energy smearing matrices for the OPERA detector in the CNGS beam.}
\end{figure}

The overall agreement in the structure and shape of the distributions between NuSmear's Default model matrix and the complete Monte Carlo simulation's matrix indicates that NuSmear provides a reasonably accurate approximation for the OPERA detector's energy smearing response.

\subsubsection{Deconstructed by particle type}

Using an example data set of muon neutrinos incident on an Argon-40 target, we construct energy smearing matrices (illustrated in Figures 6 and 7) with NuSmear's DUNE-CDR model and Default model, deconstructing them into multiple smearing matrices according to final state particle type (as itemized in Table 1 and 2 of Section 4.1.1, respectively).

\begin{figure}[H]
\includegraphics[width = 290pt]{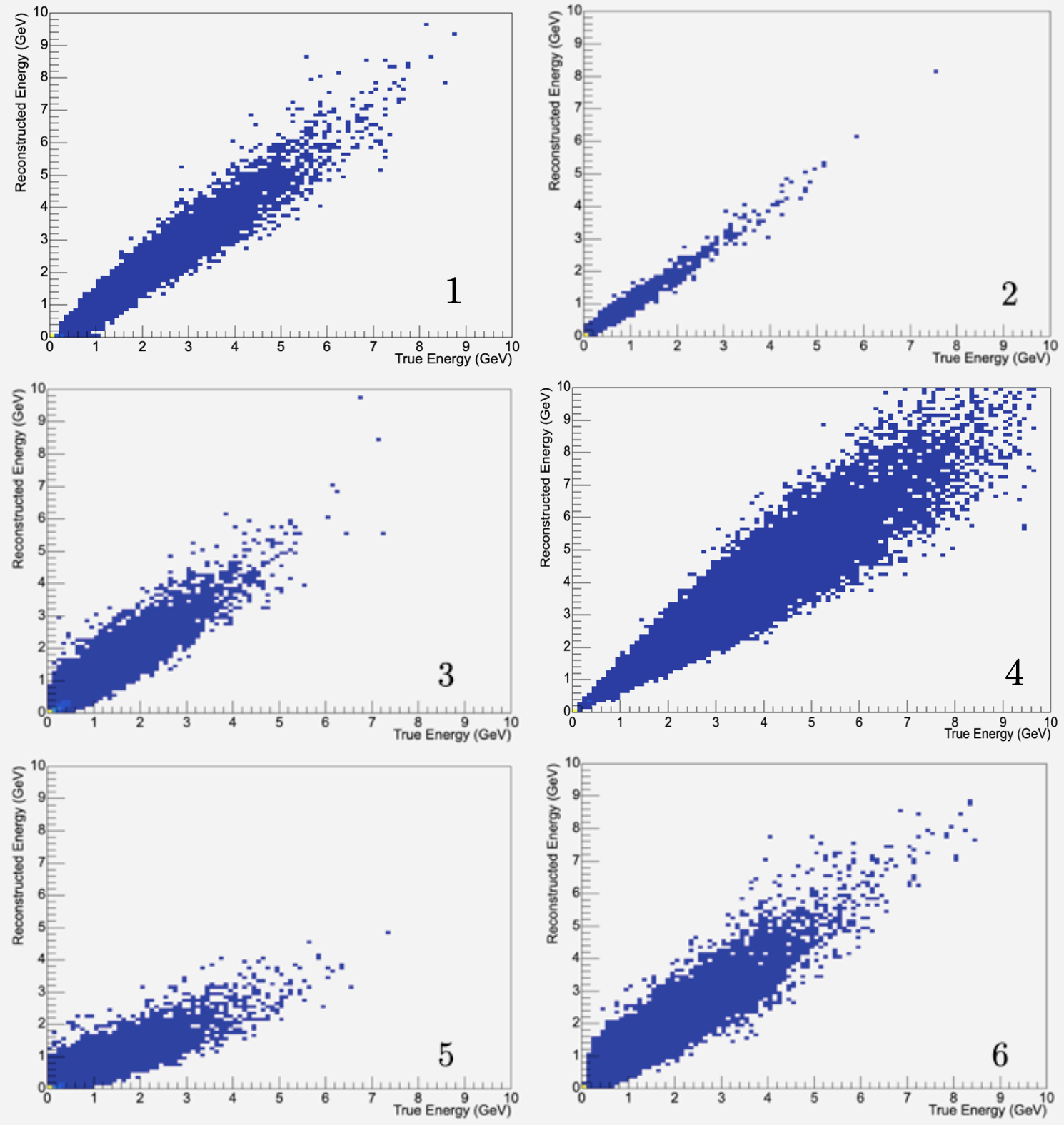}
\centering
\caption{NuSmear DUNE-CDR energy smearing matrices deconstructed by final state particle type for muon neutrinos on an Argon-40 target – $\pi^{\pm}$ (1); $\gamma$, $e^{\pm}$ (2); $p$ (3); $\mu^{\pm}$ (4); $n$ (5); other (6).}
\end{figure}

The DUNE-CDR energy smearing matrices portray significant agreement with the resolution functions and particle detection dependencies of the model. Photons, electrons, and positrons, for example, appear smeared to a lesser extent due to their on average smaller resolutions, while protons appear smeared to a greater extent due to their on average greater resolutions. Moreover, in producing the reconstructed energies of neutrons, the true energies appear to have been reduced by a constant multiplicative factor – a result which follows from the fact that only 60\% of smeared neutron energy is finally reconstructed in the DUNE-CDR model.

\begin{figure}[H]
\includegraphics[width = 290pt]{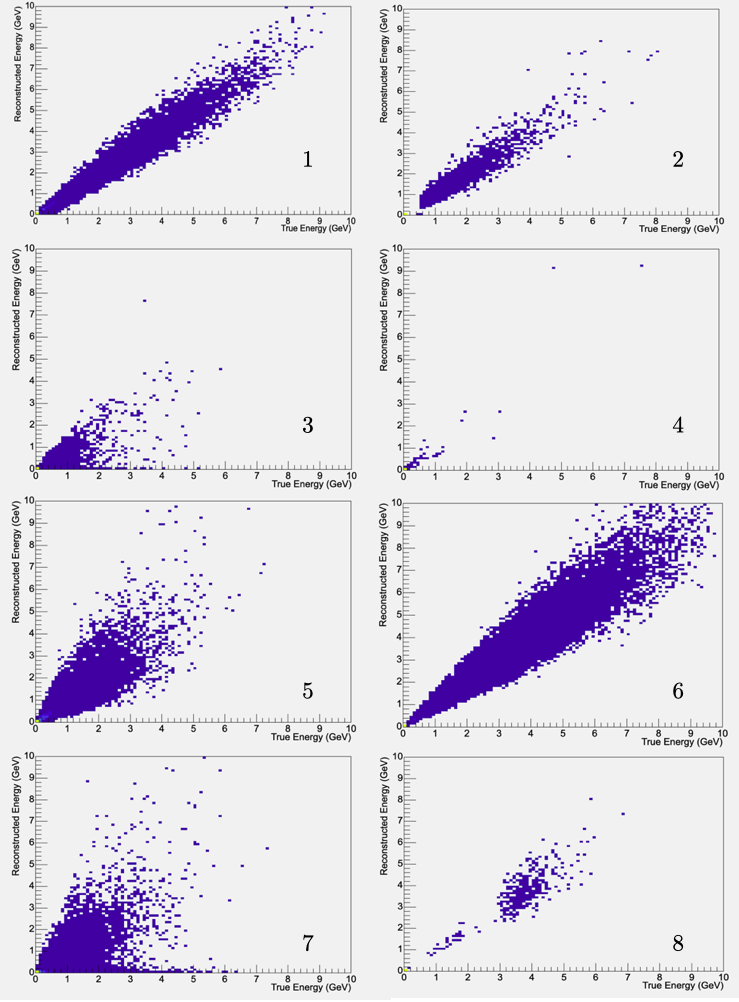}
\centering
\caption{NuSmear Default energy smearing matrices deconstructed by final state particle type for muon neutrinos on an Argon-40 target – $\pi^{\pm}$,$\pi^{0}$ (1); $K^{\pm}$, $K^{0}$/$\bar{K^{0}}$ (2); $\gamma$ (3); $e^{\pm}$ (4); $p$ (5); $\mu^{\pm}$ (6); $n$ (7); other (8).}
\end{figure}

The Default energy smearing matrices also portray significant agreement with their resolution functions and further dependencies. Particles with on average smaller energy resolutions such as pions and kaons are smeared to a comparatively lesser extent, while particles with on average greater energy resolutions such as protons and neutrons are smeared to a comparatively greater extent. Furthermore, while in the DUNE-CDR model only few data points in the neutron smearing matrix lie along the x-axis, a more significant portion of the data points in the Default model photon and neutron matrices lie along the x-axis, which predictably results from the fact that both have a 50\% chance of not being observed at all.

\subsection{Angular smearing}
\subsubsection{Complete MC comparison}

A set of events is generated using an incoming electron neutrino flux obtained from T2K experiment data, and a comparison is made between the electron angular smearing matrices constructed by NuSmear's Default model and the complete Monte Carlo simulation of the T2K Collaboration\cite{abe2020measurement,abe2011t2k}. The two matrices (see Figure 8 below), which simulate the angular smearing of the ND280 near detector, are filtered to contain only events involving charged-current (CC) interactions.

\begin{figure}[H]
\includegraphics[width = \textwidth]{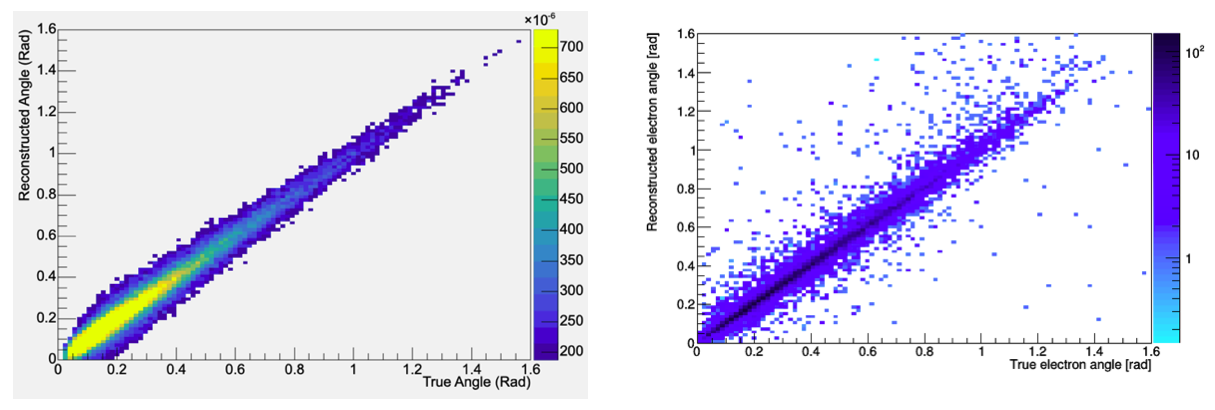}
\centering
\caption{NuSmear Default model (left) compared to complete Monte Carlo simulation (right) electron angular smearing matrices for electron neutrino CC interactions at the T2K experiment's ND280 near detector.}
\end{figure}

There is overall strong agreement between the main distributions of NuSmear's Default model matrix and the complete Monte Carlo simulation's matrix. Although some data points spread further from the main distribution are not reproduced to the same extent in NuSmear, we expect that this is a result of more complex reconstruction efficiencies and detector dependencies beyond the scope of NuSmear's geometry-independent smearing capabilities.

\subsubsection{Deconstructed by particle type}

Using an example data set of muon neutrinos incident on an Argon-40 target, we construct angular smearing matrices (illustrated in Figure 9) with both NuSmear's DUNE-CDR model and Default model, deconstructing them into multiple smearing matrices according to final state particle type (as itemized in Table 3, Section 4.1.2). Only the charged pion, proton, neutron, and muon smearing matrices are displayed below – the other angular smearing matrices can be found in the Appendix.

\begin{figure}[H]
\centering
\begin{subfigure}[H]{\linewidth}
\centering
\includegraphics[width = 290pt]{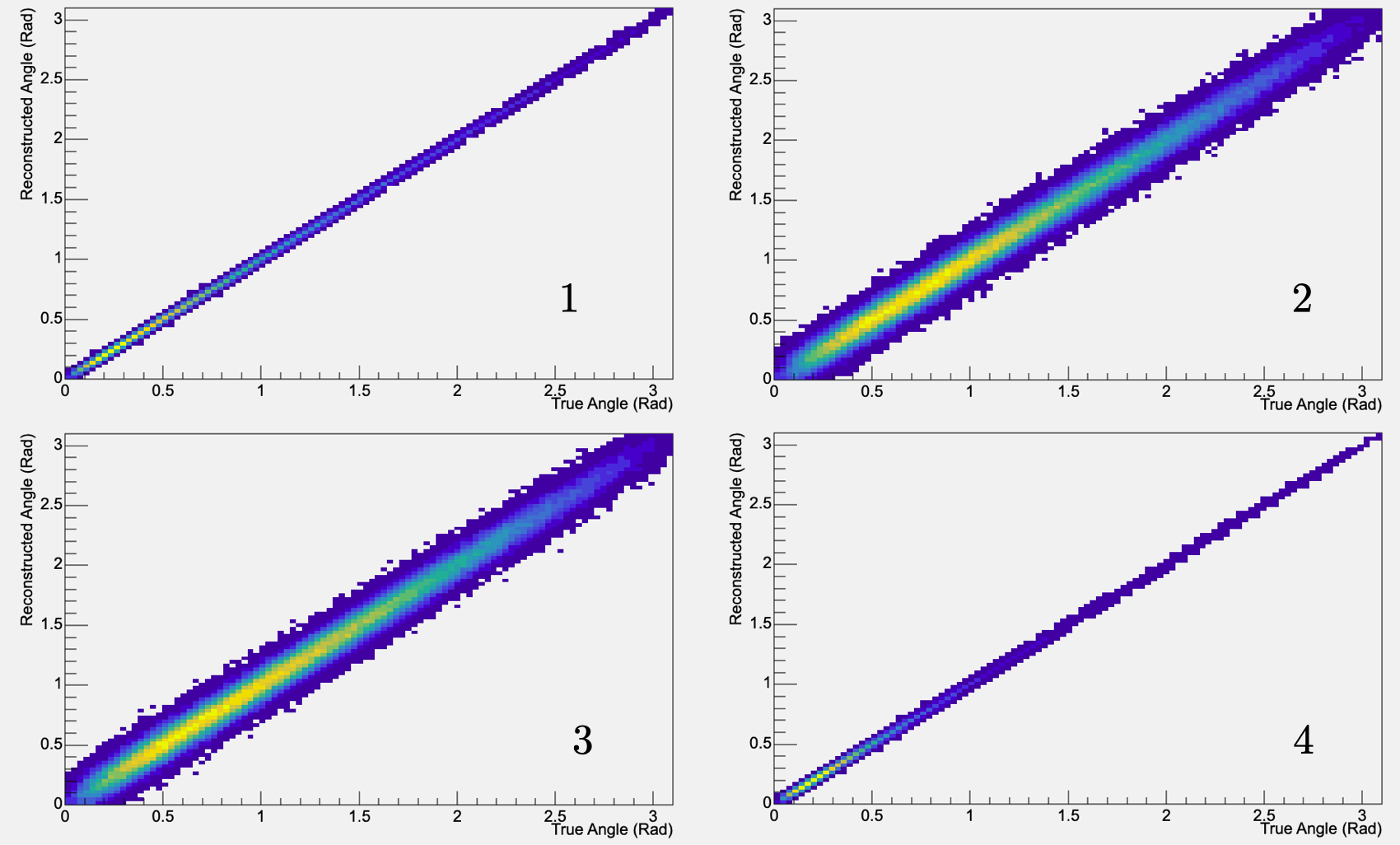}
\subcaption{DUNE-CDR angular smearing matrices}
\label{dune}
\end{subfigure}
\vskip10pt
\begin{subfigure}[H]{\linewidth}
\centering
\includegraphics[width = 290pt]{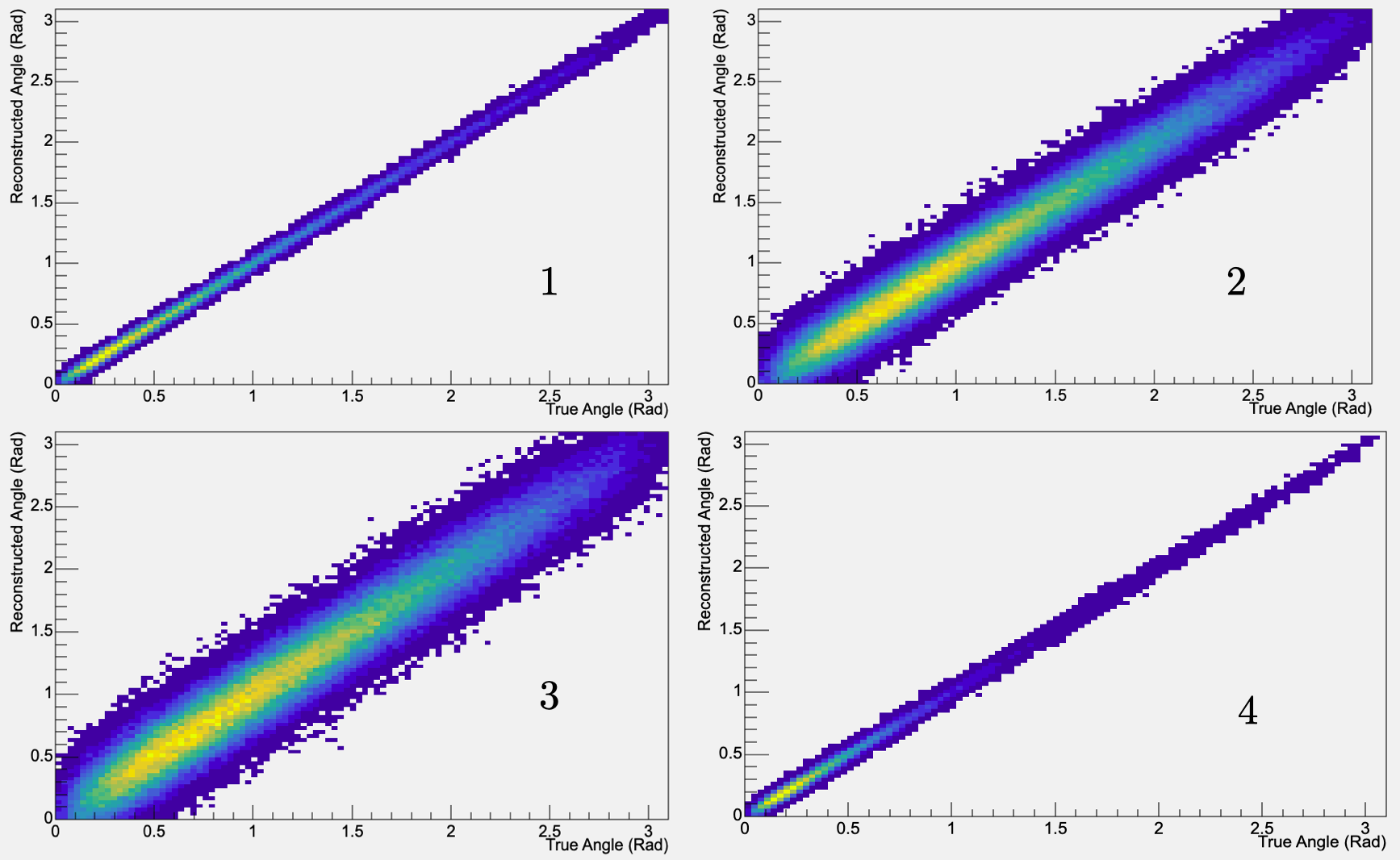}
\subcaption{Default angular smearing matrices}
\label{def}
\end{subfigure}
\caption{NuSmear DUNE-CDR and Default angular smearing matrices deconstructed by final state particle type for muon neutrinos on an Argon-40 target – $\pi^{\pm}$ (1); $p$ (2); $n$ (3);  $\mu^{\pm}$ (4).}
\end{figure}
\pagebreak

Both the DUNE-CDR and Default angular smearing matrices portray significant agreement with the resolution values of the models. The charged pion and muon matrices exhibit less smearing, reflecting their smaller resolution values, while the proton and neutron matrices exhibit more smearing, reflecting their greater resolution values. In addition, each of the Default smearing matrices exhibit a greater degree of smearing than each of the corresponding DUNE-CDR smearing matrices, reflecting the Default model's overall greater resolution values for the displayed particle types.

\section{Conclusion}

In this paper, we describe NuSmear – a generic, fast, parameterized system for simulating energy smearing and angular smearing in neutrino-nucleon scattering events. After one of NuSmear's smearing model presets (the DUNE-CDR model or the Default model) is selected, each input particle is assigned to an energy resolution function or angular resolution value, from which a particle-specific resolution is computed or obtained. NuSmear then generates a smearing distribution parameterized by the calculated resolution as well as the particle's true energy or angle, from which a reconstructed value is produced. For the angular smearing operation, the reconstructed value is returned to the user immediately, whereas in the energy smearing operation, particle detection dependencies are considered before the reconstructed value is returned to the user.

We go on to demonstrate the smearing performance of NuSmear through a series of internal and external comparisons of smearing matrices. Both NuSmear's energy smearing and angular smearing operations are validated with respect to their strong adherence to the input models, as well as their accurate reproduction of results from independent complete Monte Carlo simulations.

Hence, we conclude that NuSmear effectively satisfies the requirements of predictive accuracy and speed of computation, thereby offering itself as a robust and efficient alternative for the rapid simulation of energy and angular smearing in neutrino-nucleon scattering events.

\section{Future Prospects}

By virtue of its open access smearing software and straightforwardly adjustable smearing models, NuSmear naturally lends itself to user customization and the implementation of custom smearing configurations. Such customization could range from simply tweaking values of the numerical thresholds within resolution functions and particle detection dependencies, to applying one's own smearing models with theoretically limitless complexity. In this spirit, NuSmear users are encouraged to incorporate their own ideas into their NuSmear smearing systems, providing greater control and more precise simulation capabilities over a wide range of parameters.

\section{Acknowledgements}

This research project was undertaken as part of an internship at the University of Liverpool in the area of computational particle physics. I would like to sincerely thank my mentor, Dr. Marco Roda, for guiding me through this project and for offering helpful advice on my simulations and data analysis. I am also grateful to Professor Costas Andreopoulos for supervising my internship and providing me with the opportunity to conduct particle physics research as a high school student. Lastly, I would like to thank the Oliver Lodge Laboratory for hosting me and providing me with a comfortable environment to pursue my research.

\pagebreak
\bibliography{bibliography.bib}
\bibliographystyle{ieeetr}

\pagebreak
\section{Appendix}
\subsection{Additional NuSmear angular smearing matrices}

\begin{figure}[H]
\includegraphics[width = 290pt]{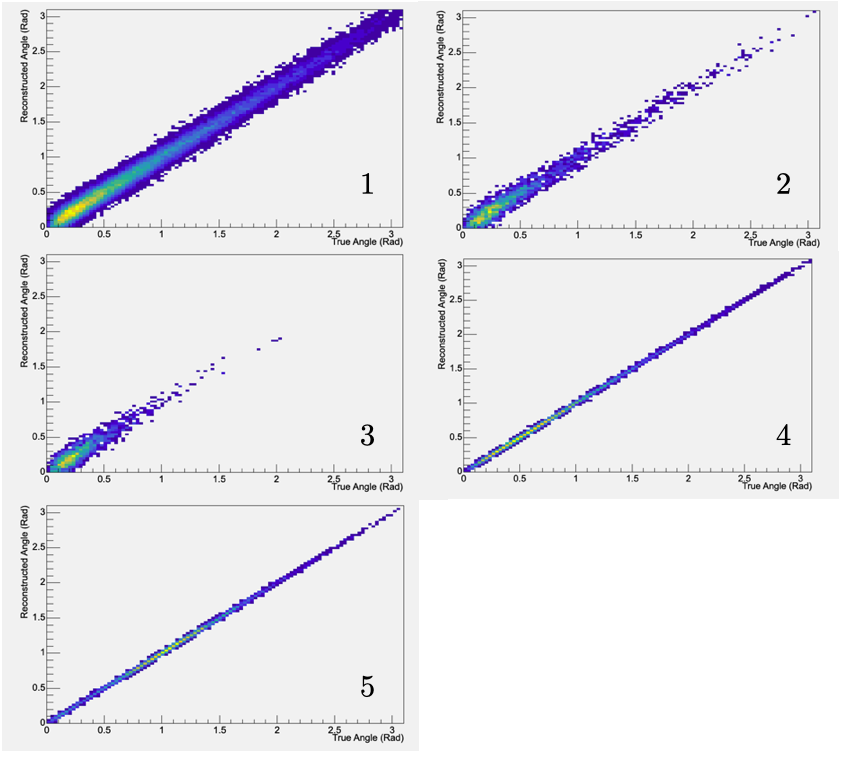}
\centering
\caption{NuSmear DUNE-CDR angular smearing matrices deconstructed by final state particle type for muon neutrinos on an Argon-40 target – $\pi^{0}$ (1); $K^{\pm}$ (2); $K^{0}/\bar{K^{0}}$ (3);  $\gamma$, $e^{\pm}$ (4); other (5).}
\end{figure}

\begin{figure}[H]
\includegraphics[width = 290pt]{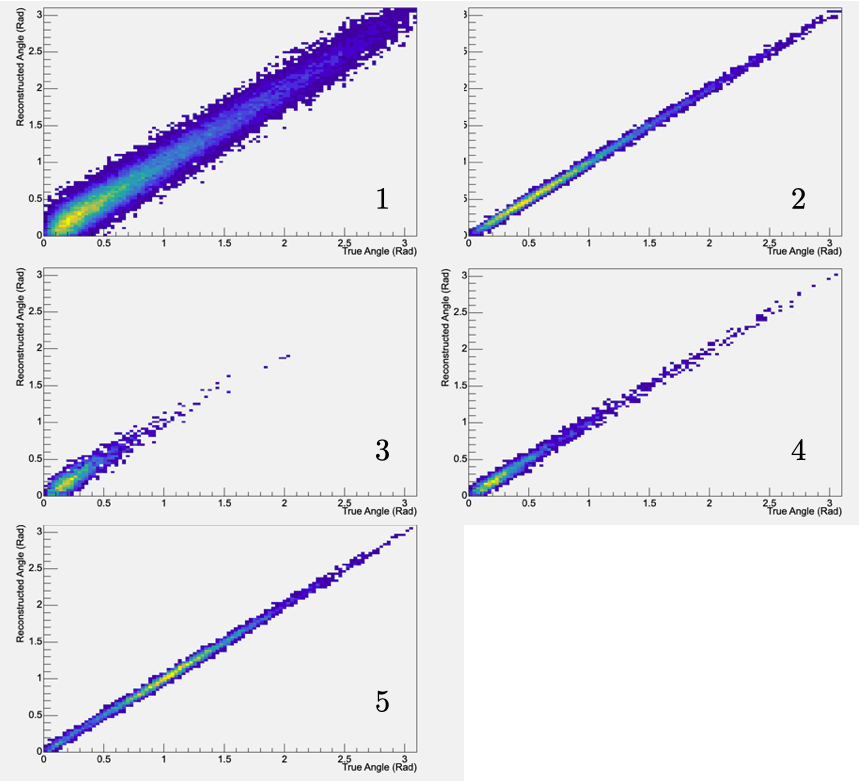}
\centering
\caption{NuSmear Default angular smearing matrices deconstructed by final state particle type for muon neutrinos on an Argon-40 target – $\pi^{0}$ (1); $K^{\pm}$ (2); $K^{0}/\bar{K^{0}}$ (3);  $\gamma$, $e^{\pm}$ (4); other (5).}
\end{figure}

\end{document}